# A new ignition hohlraum design for indirect-drive inertial confinement fusion[*]

Xin Li(李欣)[1], Chang-Shu Wu(吴畅书)[1], Zhen-Sheng Dai(戴振生)[1,†], Wu-Di Zheng(郑无敌)[1], Jian-Fa Gu(谷建法)[1], Pei-Jun Gu(古培俊)[1], Shi-Yang Zou(邹士阳)[1], Jie Liu(刘杰)[1], Shao-Ping Zhu(朱少平)[1]

[1]*Institute of Applied Physics and Computational Mathematics, Beijing 100094, China*

In this paper, a six-cylinder-port hohlraum is proposed to provide high symmetry flux on capsule. It is designed to ignite a capsule with 1.2 mm radius in indirect-drive inertial confinement fusion (ICF) . Flux symmetry and laser energy are calculated by using three dimensional view factor method and laser energy balance in hohlraums. Plasma conditions are analyzed based on the two dimensional radiation-hydrodynamic simulations. There is no $Y_{lm}$ ( $l \leq 4$ ) asymmetry in the six-cylinder-port hohlraum when the influences of laser entrance holes (LEHs) and laser spots cancel each other out with suitable target parameters. A radiation drive with 300 eV and good flux symmetry can be achieved with use of laser energy of 2.3 MJ and 500 TW peak power. According to the simulations, the electron temperature and the electron density on the wall of laser cone are high and low, respectively, which are similar to those of outer cones in the hohlraums on National Ignition Facility (NIF). And the laser intensity is also as low as those of NIF outer cones. So the backscattering due to laser plasma interaction (LPI) is considered to be negligible. The six-cyliner-port hohlraum could be superior to the traditional cylindrical hohlraum and the octahedral hohlraum in both higher symmetry and lower backscattering without supplementary technology at acceptable laser energy. It is undoubted that the hohlraum will add to the diversity of ICF approaches.

## 1. Introduction

The primary route to ignition and high gain in inertial confinement fusion (ICF) involves the use of hohlraums [1]. A hohlraum consists a high Z case with laser entrance hohles (LEHs). The laser beams are efficiently converted into x rays at the beam spots on the hohlraum wall. The study and design of a hohlraum target are essentially important in inertial fusion because it seriously influences the capsule symmetry and the hohlraum energetic, which are the most important issues of inertial fusion [1-3]. Typical capsule convergence ratios are 25 to 45, so that drive asymmetry can be no more than 1% [1], a demanding specification on hohlraum design. The flux symmetry is strongly depended on hohlraum geometries and laser beam arrangements. Up to now, various designs with different hohlraum geometries and beam arrangements have been proposed and investigated, such as cylindrical hohlraums [1,3] and spherical hohlraums with 4 LEHs [4,5] or 6 LEHs [6-8]. The cylindrical hohlraums are used most often in inertial fusion studies and are chosen as the ignition hohlraums on the US National Ignition Facility (NIF) [3]. In cylindrical hohlraums, the Legendre polynomial modes $P_2$ and $P_4$ of the flux on capsule are the main asymmetry modes required to be controlled. The $P_2$ asymmetry is controlled by using two rings per side in the hohlraum and by adjusting the power ratio between the two rings (the approach called as beam phasing technology ) on NIF [9]. However, the laser beams injected at inner cones suffer with complicated LPI issues [1], which reflect laser energy out of hohlraum directly and reduce the effect of the beam phasing technology. A crossed-beam energy transfer (CBET) [10] technique is used to maintain the required symmetry. It makes asymmetry control on NIF more complicated. The recent work [11] on NIF observed fusion fuel gains exceed unity by using a high-foot implosion method to reduce instability in the implosion. The hot spot at band-time is still far from a sphere due to the neutron image, and the analysis showed that the P2 symmetry of capsule is as high as -34%. This result shows that the capsule asymmetry is a still serious issue, which must be solved in future.

[*]Project supported by the National Natural Science Foundation of China (Grants Nos. 11435011 and 11575034)
[†]Corresponding author. E-mail: dai_zhensheng@iapcm.ac.cn

The spherical hohlraums with 4 LEHs [4,5] have no $Y_{2m}$ spherical harmonic asymmetry, but there exits $Y_{3m}$ asymmetry and it can't be eliminated by adjusting the target parameters. In the spherical hohlraums [6-8] with 6 LEHs of octahedral symmetry (octahedral hohlraums), the three dimensional laser arrangement shows that some laser beams are likely to be blocked by the high-Z plasmas created by other laser spots, which are just above these beams, destroying the symmetry control scheme. And some laser spots are very close to their neighbor LEHs in those hohlraums. Considering the nominal beam pointing errors, these laser beams are likely to transfer outside hohlraum directly. The high laser intensity of the octahedral hohlraums, due to the small laser spot, can result in high LPI risk, since LPI liner gains are proportional to the laser intensity [1]. And the laser beams are likely to be absorbed by the plasma on LEH due to the very small LEH radius. Besides, when the LEHs are not placed at the specific hohlraum-to-capsule radius ratio of 5.14 [6], there is a residual $Y_{4m}$ asymmetry and it is difficult to eliminate this asymmetry by adjusting target parameters.

In this paper, we investigate a hohlraum with six cylinder ports from theoretical side, addressing the most important issues of the flux symmetry ,laser energy and LPI.

2. **Target study**

We consider a six-cylinder-port hohlraum with 192 laser beams for a capsule of 1.2 mm radius which is designed for a 300 eV radiation drive. The details of capsule are not discussed here. The 192 beams are clustered in 48 quads of four beams [1,3]. Eight laser quads entering each LEH is in one cone at $\theta_L = 55^0$ as the opening angle that the laser quad beam makes with the LEH normal direction. The laser power is shaped to meet the requirement of capsule with peak power of 500 TW and total laser energy of 2.3 MJ. Each quad is focused to an elliptical spot, which reduces laser intensity and the long axis of the ellipse is chosen so there is no loss of LEH clearance. The nominal spot is $400 \times 600$ μm at best focus. The LEH size is designed to prevent laser absorption in the plasma of LEH edges. We choose the LEH radius at $R_{LEH} = 1400$ μm , which is much larger than that in the octahedral hohlraums [6-8]. So our hohlraum has lower laser entering risk than the octahedral hohlraums. The hohlraum material is gold and the hohlraum is filled with helium gas at density 1.5 mg/cc which is confined by a window over the LEH. The gas tamps the motion of the wall and helps to improve laser propagation [1].

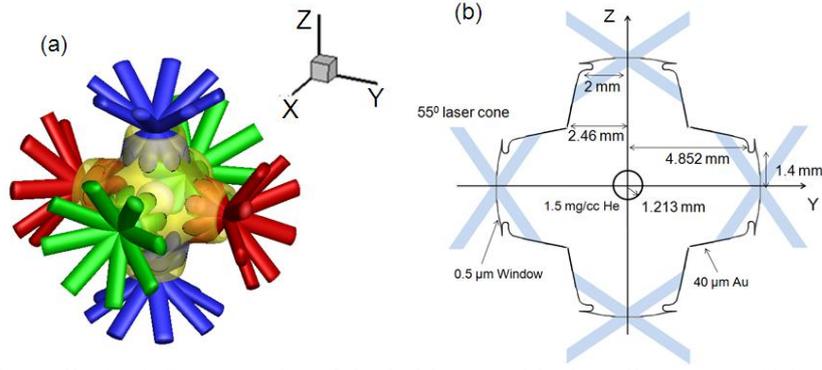

**Fig. 1.**(color online) (a) Scenography of the hohlraum with six cyliner ports, 48 laser quads and centrally located fusion capsule. (b) Hohlraum specifications.

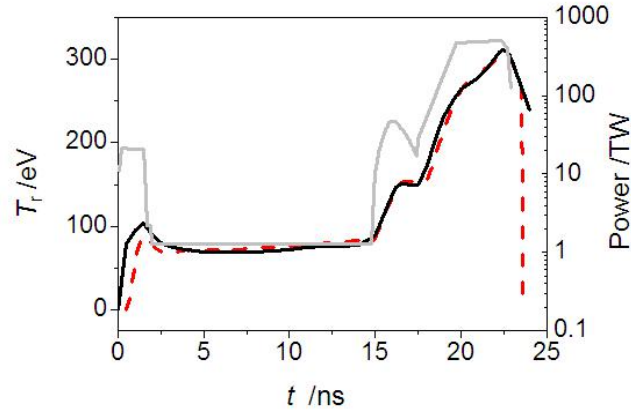

**Fig. 2.**(color online) Laser power to drive the target(grey, right scale), the resulting temperature in the hohlraum from simulations(black, left scale) and the temperature required by the capsule with 1.2 mm radius(red, left scale)

As described in Ref. [12], there is an optimal configuration of $N_L$ LEHs on an outer sphere with radius $R_{outer}$, in which the total quantity of the $N_L$ LEHs $a_l^{LEH,total}$, from $l=1$ to $l=l_0$ ( $l_0=3$ when $N_L=6$ ), are all zero, and $a_{l_0+1}^{LEH,total} \neq 0$. $a_l = \sqrt{\sum_{m=-l}^{l} |a_{l,m}|^2}$ is chosen to describe the $l$-order flux asymmetry, where $a_{l,m}$ is spherical harmonic decomposition of the flux after removing a "closed total hohlraum wall" flux [12]. Because the smooth factor vanished quickly for large $l$, $a_{l_0+1}^{LEH,total}$ dominates the asymmetry modes for $l>l_0$. Ref. [12] also points that a ring of $N_s$ identical spots incident from a LEH is equivalent with a LEH when $N_s$ is greater than some value ( $N_s \geq 5$ when $N_L=6$ ). In our design, the $Y_{lm}$ ( $l=1,2,3$ ) asymmety contributed by the LEHs and the laser spots are all zero since $N_L=6$ and $N_s=8$. From Eq. (17) in Ref. [12], the 4-order contribution $a_4^{spot,total}$ of laser spots, which is proportion to $P_l(\cos 2\theta_s)$, could be zero at the nodes of $P_4(\cos 2\theta_s)$ at $\theta_s = 15.28^0, 35.06^0, 54.94^0,$ and $74.72^0$. $\theta_s$ is the incident angel that the laser beam makes with the LEH normal direction on an outer sphere with radius $R_{outer}$ and $P_l(x)$ is the $l$-order Legendre polynomial function. $a_4^{spot,total}$ is positive when $\theta_s > 74.72^0$. The same equation also shows that the LEH contribution $a_4^{LEH,total} < 0$. The negative sources of LEH and the positive sources of laser spot with $\theta_s > 74.72^0$ can cancel each other out to make $a_4^{total} = 0$ ( $a_4^{total} = a_4^{spot,total} + a_4^{LEH,total}$ ). In the six-cylinder-port hohlraum (Fig. 3), the radius $R_{outer,spot}$ of the outer sphere with all laser spots on is smaller than the radius $R_{outer,LEH}$ of the sphere with all LEHs on. So although the laser incident angle $\theta_L$ is $55^0$, the equivalent laser incident angle $\theta_s$ in sphere with $R_{outer,spot}$ is about $80^0$. It makes the cancelling possible. Instead, in the octahedral hohlraums [6-8], if $\theta_s > 74.72^0$, the laser beams will be very close to the hohlraum wall near LEHs and be prone to be absorbed by the wall blowoff plasma near LEHs.

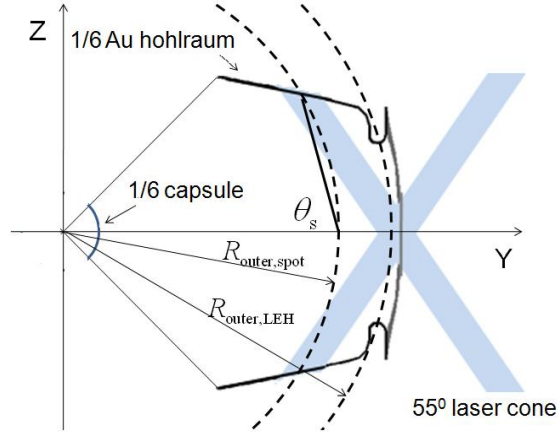

**Fig. 3.** Scenography of one single port

To certify the above theoretical prediction, we further use a three-dimensional view factor code, based on Ref. 13, to calculate the radiation flux on the capsule. The asymmetry of flux on capsule can be decomposed into the spherical harmonics defined in quantum mechanics and $b_{lm}$ is defined as harmonic decomposition. We further define $C_{lm} = b_{lm}/b_{00}$ to describe the quantity of the $Y_{lm}$ asymmetry. Shown in Fig. 4 is the flux distribution on the capsule of 1.2 mm radius. Although the LEHs with $R_{outer,LEH}/R_C = 4$ are not at the node of $Y_{4m}$ smooth factor, we can obtain $C_{4m} \sim 0$. We adjust the length of port, the radius of port, the incident angel of laser quad, and the position of laser ring to make the cancelling of spots and LEHs happen. Of course, while a large number of target parameters are adjusted to meet the symmetry requirement of capsule, the laser energy should be acceptable during the whole adjusting procedure. According to Ref. 12, the $Y_{lm}$ ( $l = 1, 2, 3$ ) asymmety contributed by the LEHs and the laser spots are all absolutely zero during the whole laser duration because the LEHs and the laser spots maintain octahedral symmetry over time, respectively. $Y_{4m}$ is the only remaining asymmetry, which varies with time because the positions of laser spots change over time due to the wall motion. Just like cylindrical hohlraums [1,3], we can only adjust time-integrated $Y_{4m}$ asymmetry in six cylinder port hohlraums mainly by changing the initial positions of the laser spots. In our target design, the laser spots move to the LEHs about $300\,\mu m$ by analyzing the simulation results (see below) during the whole laser duration. Considering the motion of the laser spots, the three dimensional view factor calculations show that the $Y_{4m}$ asymmetry varies from 1% to 0 (about 1% at the first two nanoseconds and about zero in the main pulse). And the values of $Y_{4m}$ are much better than the symmetry demands of our ignition capsule (The details of the capsule will be discussed in another paper).

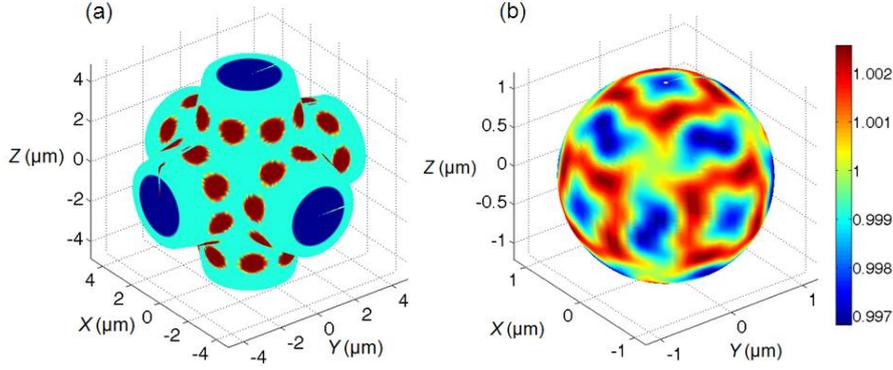

**Fig. 4.** (color online) (a) Spot pattern on the wall of the six-cylinder-port hohlraum. (b) The relative flux ($F/\bar{F}$) distribution on the capsule with 1.2 mm radius. $\bar{F}$ is the averaged flux.

We use the energy balance [1] to relate the internal hohlraum radiation temperature to the input laser energy by balancing the absorbed laser energy with the x-ray energy radiated into the wall, $E_W$, absorbed by the capsule, $E_C$, and the energy that escapes through the LEH, $E_{LEH}$,

$$\eta_{CE}\eta_a E_L = E_W + E_C + E_{LEH} \qquad (1)$$

where $\eta_a$ is the absorbed laser efficiency and $\eta_{CE}$ is the x-ray conversion efficiency from laser energy to soft x-rays. $E_W \propto \tau^{0.56} T_r^{3.45} A_W$, $E_C \propto \tau T_r^4 A_C$, and $E_{LEH} \propto \tau T_r^4 A_{LEH}$. $\tau$ is the main pulse duration. $T_r$ is the peak radiation temperature. $A_W$ is hohlraum wall area, $A_C$ is capsule area, and $A_{LEH}$ is LEH area. Usually, $\eta_{CE}$ is around $90\%$ on NIF [8,14]. NIF experimental results [15] of cylindrical hohlraums shows that $\eta_a$ is about $85\%$. The $15\%$ losses are almost from the inner cones due to Simulated Raman Scattering (SRS) during peak laser power and the outer cones have little energy losses. In our target design, each port has one cone and its incident angle, laser intensity and plasma conditions are all similar to those of outer cones of cylindrical hohlraums on NIF. So $\eta_a$ of hohlraum with 6 cylinder ports should be higher than cylindrical hohlraums. In this study, $\eta_a \sim 1$. We apply this energy balance to the ignition target hohlraum with six cylinder ports and find that 2.3 MJ laser energy is required to produce 300 eV peak radiation temperature.

To study the plasma conditions of the cylinder ports by simulations, we consider one cylinder port with $1/6$ laser energy on one cone in cylindrical symmetry, because we do not have a three dimensional hydrodynamics code for the six-cylinder-port hohlraums. $1/6$ capsule is inside the cylinder port. This simulation scheme is reasonable since the six cylinder ports are relatively independent on each other. It is noticeable that the sum of six single cylinder ports is not equivalent to the hohlraum with six cylinder ports, but the differences are too small to influence the plasma conditions and the energy balance very much. We use a two dimensional non-equilibrium radiation hydrodynamics code LARED-Integration [16]. Capsule needs a shaped radiation temperature to launching sequential shocks [1]. In our design, laser power is tuned to meet the driven temperature requirement of the capsule with 1.213 mm radius. Shown in Fig. 2 is the laser power put into the port and the radiation temperature calculated by simulations. From simulations, the peak power and the total laser energy of the sum of six single ports are $500\,\text{TW}$ and $2.4\,\text{MJ}$, which is close to the energy estimated by energy balance.

Maps of the spatial distribution of the electron density, the electron temperature, the x-ray emission, and the laser absorption at peak power are shown in Fig. 5. The laser beams (with laser intensity at $I_L \sim 9.5 \times 10^{14}\,\text{W/cm}^2$) impact the Au hohlraum wall and creates a Au plasma with high electron temperature ($T_e \sim 6.5\,\text{keV}$) and low density ($n_e \sim 0.09 n_c$). $n_c$ is the laser critical density. For a laser wavelength $\lambda$ in microns, $n_c(\text{cm}^{-3}) = 1.1 \times 10^{21}/\lambda(\mu m)^2$. The electron temperature, the electron density, the plasma length ($L$) and the laser intensity on laser cone are very similar to those on outer cones of cylindrical hohlraums[17] at $300\,\text{eV}$ of radiation temperature. The linear gain[18] of Simulated Brillouin Scattering (SBS) is proportional to $I_L L n_e / T_e$. Analysis[17] shows that SBS rising from Au plasma is the main LPI risk on outer cones of cylindrical hohlraums on NIF. Since experiment results[15] on NIF do not show obvious SBS on outer cones of cylindrical hohlraums, there should be little SBS in the laser beams of the six-cylinder-port hohlraum. That is why we suppose $\eta_a \sim 1$ before.

Without the gas fill, for laser pulses as long as those required for ignition, the Au blowoff driven by laser and radiation has time to fill the hohlraum, and the laser beam will not propagate to the initial impacting point since it is absorbed by the Au plasma filling in the hohlraum. This changes the x-ray emission position and destroys the symmetry control scheme. In our design, He gas at $1.5\,\text{mg/cc}$ initial density holds back the Au blowoff and ensures that the path of laser cone is clear (Fig. 6(c)) and the x-ray emission position is very close to the initial laser impacting points (Fig. 6(d)). This is the reason that we can use the view factor code to study the flux asymmetry on capsule.

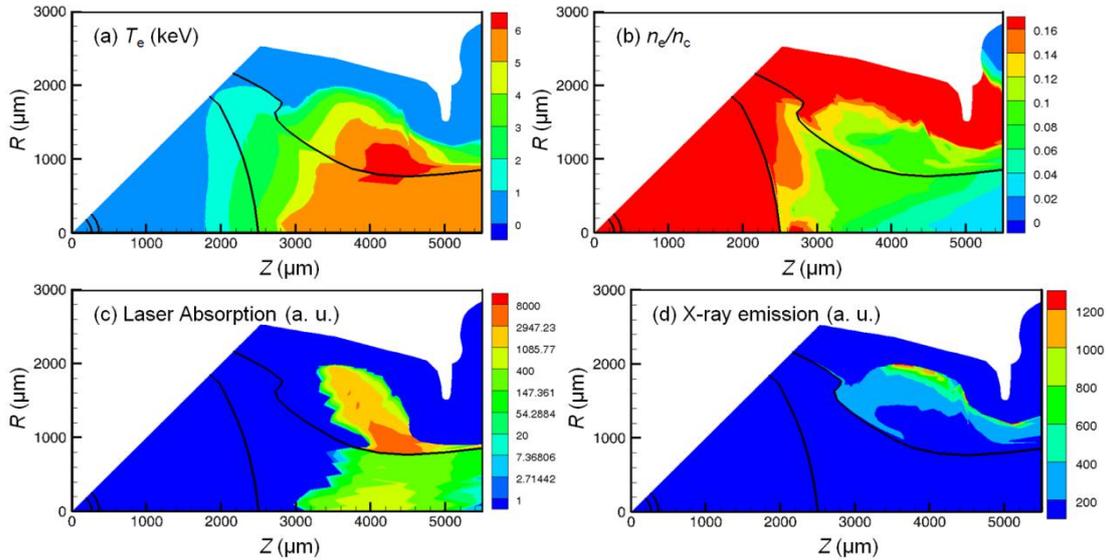

**Fig. 5.** (color online) Maps of plasma at peak power: (a) $T_e$ in keV. (b) $n_e/n_c$. (c) Laser asorption. (d) x-ray emission

## 2.1. Conclusions

In summary, we have proposed a new ignition hohlraum with six cylinder ports for the capsule with $1.2\,\text{mm}$ radius. We use the three dimensional view factor calculations, the energy balance analysis, and the two dimensional hydrodynamic simulations to study the hohlraum. Our theoretical study shows that the six-cylinder-port hohlraum has very high flux symmetry on capsule. Specially, the $Y_{4m}$ asymmetry, which exists in the octahedral hohlraums, can be adjusted to zero by cancelling the influences of laser spots and LEHs each other out, even though the LEHs are not on a sphere with golden radius ratio. About 2.3 MJ laser energy is required to provide the radiation driven of capsule. The simulations showed that the plasma conditions in the hohlraum should simulate little SBS and the x-ray emission positions are still very close to the initial ones. Obviously, in addition to the residual $Y_{4m}$ problem, the other problems about laser transferring and LPI, mentioned before in the octahedral hohlraums, have been solved or eased in the six-cylinder-port hohlraum. Next, we plan to study the time-varing $Y_{4m}$ asymmetry, the asymmetry of M-band flux and the laser beam entering problem more deeply because there is a larger optimization space in the target design. Of course, in future, three dimensional simulations are necessary to classify the details of asymmetry, plasma, and efficiency in hohlraums. And aboundant experiments are worthy to be done with six-cylinder-port hohlraums on exiting facilities, since we have little experience on laser arrangement, target fabrication, and diagnostics.

**Acknowledgment**

The The authors wish to acknowledge the beneficial discussions with Hao Duan(段浩), Hua-Sen Zhang(张桦森), Yi-Qing Zhao(赵益清) and Ke Lan(蓝可)

**References**
1. Lindl J D, Amendt P, Berger R L, Glendinning S G, Glenzer S H, Haan S W, Kauffman R L, Landen O L and Suter L J 2004 *Phys. Plasmas* 2, 3933
2. Atzeni S and Meyer-ter-Vehn J 2004 *The Physics of Inertial Fusion* (Oxford Science, Oxford)
3. Haan S W, Lindl J D, Callanhan D A, Clark D S, Salmonson J D, Hammel B A, Atherton L J, Cook R C, Edwards M J, Glenzer S, Hamza A V, Hatchett S P, Herrmann M C, Hinkel D E, Ho D D, Huang H, Jones O S, Kline J, Kyrala G, Lanen O L, MacGowan B J, Marinak M M, Meyerhofer D D, Milovich J L, Moreno K A, Moses E I, Munro D H, Nikroo A, Olson R E, Peterson K, Pollaine S M, Ralph J E, Robey H F, Spears B K, Springer P T, Suter L J, Thomas C A, Town R P, Vesey R, Weber S V, Wilkens H L, and Wilson D C 2011 *Phys. Plasmas* 18 051001
4. Phillion D W and Pollaine S M 1994 *Pyhs. Plasmas* 1 2963
5. Schnittman J D and Craxton R S 1996 *Pyhs. Plasmas* 3 3786
6. Lan K, Liu J, Lai D X, Zheng W D and He X T 2014 *Pyhs. Plasmas* 21 010704
7. Lan K, He X T, Liu J, Zheng W D and Lai D X 2014 *Pyhs. Plasmas* 21 052704
8. Lan K and Zheng W D 2014 *Phys. Plasmas* 21 090704
9. Kyrala G A, Kline J L, Dixit S, Glenzer S, Kalantar D, Bradley D, Lzumi N, Meezan N, Landen O, Callahan D, Weber S V, Holder J P, Glenn S, Edwards M J, Koch J, Suter L J, Haan S W, Town R P J, Michel P, Jones O, Langer S, Moody J D, Dewald E L, Ma T, Ralph J, Hamza A, Dzenitis E and Kilkenny J 2011 *Phys. Plasmas* 18 056307
10. Michel P, Glenzer S H, Divol L, Bradley D K, Callahan D, Dixit S, Glenn S, Hinkel D, Kirkwood R K, Kline J L, Kruer W L, Kyrala G A, LePage S, Meezan N B, Town R, Widmann K, Williams E A, MacGowan B J, Lindl J and Suter L J 2010 *Phys. Plasmas* 17 056305

11. Hurricane O A, Callahan D A, Casey D T, Celliers P M, Cerjan C, Dewald E L, Dittrich T R, Doppnerl T, Hinkel D E, Berzak Hopkins L F, Klines J L, LePage S, Ma T, MacPhee A G, Milovich J L, Pak A, Park H -S, Patel P K, Remington B A, Salmonson J D, Springer P T and Tommasini R 2014 *Nature* 506 343
12. Duan H, Changshu Wu, Wenbing Pei and Shiyang Zou 2015 *Phys. Plasmas* 22 092704
13. Cohen D H, Landen O L and MacFarlane J J 2005 *Phys. Plasmas* 12 122703
14. Olson R E, Suter L J, Kline J L, Callahan D A and Rosen M D 2012 *Phys. Plasma.* 19 053301
15. Lindl J, Landen O, Edwards J, Moses E and NIC Team 2014 *Phys. Plasmas* 21 020501
16. Song P, Zhai C L, Li S G, Yong H, Qi J, Hang X D, Yang R, Cheng J, Zeng Q H, Hu X Y, Wang S, Shi Y, Zheng W D, Gu P J, Zou S Y, Li X, Zhao Y Q, Zhang H S, Zhang A Q, An H B, Li J H, Pei W B and Zhu S P 2015 *High Power Laser and Particle Beams* 27(3) 032007
17. Hinkel D E, Haan S W, Langdon A B, Dittrich T R, Still C H and Marnak M M 2003 *Phys. Plasmas* 11(3) 1128
18. Laffite S and Loiseau P 2010 *Phys. Plasmas* 17 102704